\documentclass[journal, twocolumn, final]{IEEEtran}

\usepackage[utf8]{inputenc}
\usepackage{amsmath}
\usepackage{amsfonts}
\usepackage{amssymb}
\usepackage{amsthm}
\usepackage{bm}
\usepackage{enumerate}
\usepackage{algorithm}
\usepackage{algorithmicx}
\usepackage{algpseudocode}

\usepackage{xspace}
\usepackage{color}

\usepackage{balance}

\usepackage{url}
\usepackage{hyperref}
\usepackage[capitalise,noabbrev]{cleveref}
\usepackage{cite}

\usepackage{graphicx}
\usepackage{placeins}
\usepackage [english]{babel}
\usepackage [autostyle, english = american]{csquotes}
\usepackage{authblk}
\title{Optical Proof of Work}
\author[1]{Michael Dubrovsky}
\author[2]{Marshall Ball}
\author[1]{Bogdan Penkovsky}
\affil[1]{PoWx, Cambridge, MA, USA. Email: mike@PoWx.org, bogdan@PoWx.org}
\affil[2]{Columbia University, New York, NY, USA. Email: marshall@cs.columbia.edu}

\begin{document}

\maketitle

\thispagestyle{plain}
\pagestyle{plain}

\begin{abstract}
    \noindent 
Most cryptocurrencies rely on Proof-of-Work (PoW) ``mining'' for resistance to Sybil and double-spending attacks, as well as a mechanism for currency issuance. SHA256-based PoW (Hashcash) has successfully secured the Bitcoin network since its inception, however, as the network has expanded to take on additional value storage and transaction volume, Bitcoin PoW's heavy reliance on electricity has created scalability issues, environmental concerns, and systemic risks. Mining efforts have concentrated in areas with low electricity costs, thus creating single points of failure. Although the security properties of PoW rely on imposing a trivially verifiable economic cost on miners, there is no fundamental reason for it to consist primarily of electricity cost. To scale systems like Bitcoin to 10-100x its current size, the authors propose a novel PoW algorithm, Optical Proof of Work (oPoW), to eliminate energy as the primary cost of mining. Optical Proof of Work imposes economic difficulty on the miners, however, the cost is concentrated in hardware (capital expense—CAPEX) rather than electricity (operating expenses—OPEX). The oPoW scheme involves minimal modifications to Hashcash-like PoW schemes and thus inherits many properties from such schemes, including basic safety/security from SHA or a similar hash function. 

Rapid growth and improvement in silicon photonics over the last two decades has recently led to the commercialization of silicon photonic co-processors (which are based on integrated circuits that use photons instead of electrons to perform specialized computing tasks) for low-energy deep learning computations. oPoW is optimized for a simplified version of this technology such that miners are incentivized to use specialized, highly energy-efficient photonics for computation.

Beyond providing energy savings, oPoW has the potential to improve network scalability, enable decentralized mining outside of low electricity cost areas, and democratize issuance. Geographic decentralization will make the oPoW mining ecosystem more censorship-resistant, with reduced exposure to partition attacks and regional regulations. Additionally, due to the CAPEX dominance of mining costs, oPoW hashrate will be significantly less sensitive to underlying coin price declines. In this paper, we provide an overview of the oPoW concept, algorithm, and hardware. 

\end{abstract}
\begin{figure}[]
    \centering
\includegraphics[width=0.46\textwidth]{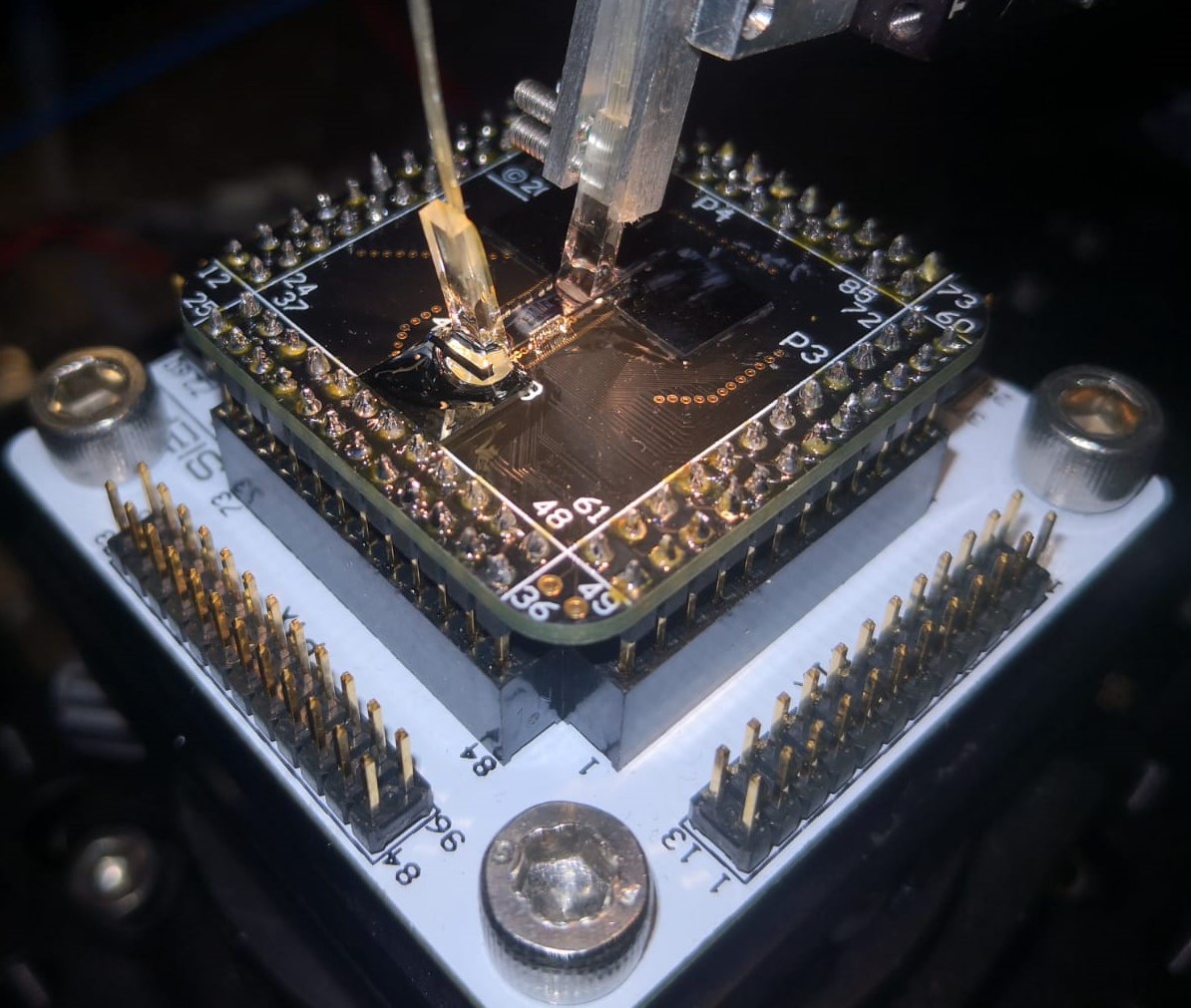}
  \caption{oPoW Silicon Photonic Miner Prototype}
  \label{fig:chip_carrierw}
\end{figure}

\section{Introduction}

The primary function of public cryptocurrency networks, such as Bitcoin, is to maintain a decentralized electronic ledger of transactions. Crucially, this requires that there be no single authority, such as a bank, controlling or validating the contents of the ledger. A naive design of this kind of network may be achieved if users post their transactions publicly via signed messages (using public-key cryptography), and a transaction is considered complete only when the majority of nodes on the network have accepted it. 
However, if the network is to be trustless/permissionless and resilient to malicious actors, a mechanism must exist to prevent
Sybil attacks\footnote{Multiple nodes controlled by one malicious actor} and double-spending\footnote{Making two purchases with the same coin by rewriting the ledger to remove the first transaction}. 

Although there were past attempts at e-cash systems, Bitcoin's architecture (outlined in the original Nakamoto whitepaper \cite{Bitcoin}) was the first to
solve the double-spend and Sybil attack problems through clever use of Hashcash \cite{hashcash} \textit{Proof of Work} (PoW). Nakamoto's key insight was that Proofs of Work enable distributed systems to automatically impose trivially verifiable costs on participating nodes, allowing for byzantine agreement \cite{Byzantine} in settings previously believed to be intractable, with the added feature of creating a distribution mechanism for the cryptocurrency, i.e. nodes are compensated for their ``work" using the cryptocurrency itself. 
This has led to the incredible rise of Bitcoin and other cryptocurrencies, creating the possibility of unprecedented access to financial freedom and property rights globally.

\subsection{Proof of Work in the Context of Blockchains}

Proof of Work schemes, or pricing functions, were initially proposed at Crypto 1992 by Cynthia Dwork and Moni Naor for a variety of tasks such as combating junk mail \cite{Dwork_Naor}. A ``Proof of Work" is a solution to a specific computing challenge that unavoidably requires a certain amount of computational work to solve. This challenge is called a cryptopuzzle and is designed in such a way that it can only be brute forced by checking all possible solutions one by one until a valid solution is found. This assures that solutions are relatively rare. Solving these cryptopuzzles in the context of cryptocurrency is colloquially known as \textit{mining}, because a successful solution yields rewards (known as ``block rewards"). On the other hand, it is easy to verify any solution is correct, requiring only a single cycle of computation. 
Therefore, a Proof of Work provides a trivially verifiable guarantee that a certain amount of computation was performed to produce it. 

In Bitcoin, transactions are recorded into \textit{blocks}, and a linked sequence of such blocks is called a \textit{block chain}. Once a mining device (\textit{miner}) compiles a block of transactions, it shuffles through random values of a special input (\textit{nonce}) in the block until a cryptographic hash of the block is smaller than a predetermined threshold. The security properties of hash functions force a miner to test nonces by brute force until a satisfactory block is found. Such a block constitutes a solution to the cryptopuzzle and is itself the proof of work. Once the block is published, anyone can easily verify that the work was done by calculating the cryptographic hash of the block and checking that it is really below the predetermined threshold\footnote{This threshold is automatically adjusted by the system such that only 1 block is found every $\sim 10$ minutes; the lower the threshold the more unlikely it is to find a solution. For example: A SHA256 hash produces 256 bits, if the difficulty setting requires the miner to find a nonce that leads to a block hash with 40 leading zeros, then statistically $2^{40}$ trials will be required to find one of the ``acceptable" nonces.}. Bitcoin uses the cryptographic hash SHA-256(NIST), but various hash functions are used by different blockchain networks. Each type puts different load on the processor and memory of a miner’s computing device, but they all use the same principles. Ethereum for example, uses a cryptographic hash named Ethash, which has greater memory requirements\cite{Ethereum}. 

As a result, a blockchain's validity is based on previously performed computational work. That also means that the longest chain, implicitly corresponding to the highest amount of work, can be automatically considered to be the valid transaction history (as it accrued the majority of computing resources). Modifying any single block requires a vast amount of computation\footnote{The proofs of work for all blocks following the altered block must be recomputed.}, which quickly becomes infeasible without control of more than half the computing power in the network(this is called a $51\%$ attack, see Section \ref{sec:51}). Moreover, any double-spending transaction becomes impossible as only the longest of the two newly created blockchains will be recognized as valid. PoW has also been applied to more complex high-throughput
(in transactions per second)
decentralized ledgers where blocks are in a directed acyclic graph, not simply a chain\cite{DBLP:journals/iacr/BentovHMN17,Sompolinsky2018PHANTOMG}. PoW schemes have an excellent track record of ensuring the irreversibility of transactions in the Bitcoin network. However, Proof of Work has run into severe scaling issues that may eventually undermine Bitcoin's growth. 

\subsection{Challenges Faced by Bitcoin's Proof of Work Ecosystem}

As Bitcoin has grown over the past decade from a small network run by hobbyists to a global currency, the underlying Proof of Work protocol has not been updated. Initially envisioned as a global decentralized network (``one CPU-one vote"), Bitcoin transactions today are secured by a small group of corporate entities. Due to the increase in the market value of mining rewards over time and competition between miners, Bitcoin mining difficulty has grown exponentially, leading to the industrialization of mining. The enormous and growing\footnote{Hashrate, and therefore energy use, is designed to grow with network value to maintain an unfeasible high cost of double-spending relative to the overall value of a successful double-spend attack.} energy use of Proof of Work has led to geographic centralization of mining in purpose-built data centers located in regions with very low energy costs and barred small entities from the mining ecosystem. 

Although the exact numbers are disputed, Bitcoin's energy use has grown steadily with it's market value, and, today, Bitcoin is estimated to consume over 75 terawatt-hours per year\cite{Cambridge}. Given that this is more than the electricity consumption of Austria\cite{IEA}, Bitcoin mining heavily favors economies of scale. In fact, it is only feasible for entities that can secure access to abundant, inexpensive energy\cite{Forbes}. The economics of mining limit profitability to places like Iceland and Western China. Besides the negative environmental externalities, which may be significant, mining today is performed primarily with the consent (and in many cases, partnership) of large public utilities and the governments that control them. Although this may not be a problem in the short term, in the long term it stands to erode the censorship resistance and security of Bitcoin and other public blockchains through potential regulation or partitioning attacks\footnote{Censorship at this scale is not as unlikely as one might think\cite{Russia, China}, however a lot of the risk can be mitigated by distributing mining globally.}\cite{apostolaki2017hijacking}.

An additional consequence of the energy-based economics of Proof of Work is the sensitivity of hashrate to block reward value. If the dollar value of block rewards falls or electricity prices rise, marginally profitable miners are forced to shut off their machines to avoid running at a loss. This leads to undesirable instability in the security of the network - especially during periods of volatility. 

It is important to note that, from an algorithmic standpoint, \textit{ Bitcoin's energy consumption is a feature, not a bug}. The network is designed to automatically incentivize an increase in PoW mining as it gets bigger in order to maintain a proportionally higher level of security. The dollar value of the mining rewards rises (by design) with the market value of the coin, leading miners to spend more resources competing for the mining rewards (which are denominated in Bitcoin), and therefore use more energy. This allows Bitcoin to scale the cost of a $51\%$ attack as the reward associated with a successful attack increases. The Bitcoin algorithm has no direct access to information about Bitcoin's market value, but it can indirectly infer a value increase from a hashrate increase\footnote{Although this is not perfect as hashrate tends to increase due to hardware improvements.}. An alternative mining reward algorithm can be imagined that actually reduces the block reward as hashrate increases, thus limiting the incentive for miners to expend more resources and energy\footnote{Bitcoin's block subsidy halving effectively does this to an extent, however the common argument is that transaction fees will pay for security once the block subsidy is depleted}. This would indeed decouple the energy consumption of the network from the total value stored and the market value of each coin, however, it would mean that the cost of attacking the network would no longer grow with the incentive to do so. 

New consensus mechanisms have been proposed as a means of securing cryptocurrencies whilst reducing energy cost, such as various forms of Proof of Stake\footnote{There are numerous attempted implementations, notably Ethereum's so-far unsuccessful effort to implement some form of PoS on their mainnet. These schemes also introduce new challenges \cite{DBLP:conf/cvcbt/GaziKR18}.} and Proof of Space-Time \cite{PoST,DBLP:conf/crypto/MoranO19} being implemented by Chia and Spacemesh. While many of these alternative mechanisms offer compelling guarantees, they generally require new security assumptions, which have not been stress-tested by live deployments at any adequate scale. Consequently, we still have relatively little empirical understanding of their safety. Completely changing the bitcoin paradigm is likely to introduce new unforeseen problems. We believe that the major issues discussed above can be resolved by improving rather than eliminating Bitcoin's fundamental security layer—Proof of Work.

\subsection{A Next-Generation Proof of Work}

Instead of devising a new consensus architecture to fix the scaling issues, we consider ways to shift the economics of PoW. As it is used in Bitcoin-like systems, PoW allows networks like Bitcoin to achieve consensus via economic difficulty imposed on the miners. However, the financial cost does not need to be concentrated in electricity. In fact, the situation can be significantly improved by reducing the operating expense (OPEX)—energy—as a significant cost of mining.
Then, by shifting the cost towards capital expense (CAPEX)—mining hardware—the dynamics of the mining ecosystem become much less dependent on electricity prices, and much less electricity is consumed as a whole. Moreover, this automatically leads to geographically distributed mining, as mining becomes profitable even in regions with expensive electricity. Finally, lower energy consumption eliminates heating issues experienced by today's mining operations, which further decreases operating cost as well as noise associated with fans and cooling systems. All of this means that individuals and smaller entities would be able to enter the mining ecosystem simply for the cost of a miner, without first gaining access to cheap energy and a dedicated, temperature-controlled data center. To a degree, memory-hard PoW schemes like Cuckoo Cycle \cite{Tromp}, which increase the use of SRAM in lieu of pure computation, push the CAPEX/OPEX ratio in the right direction by occupying ASIC chip area with memory.    

To maximize the CAPEX to OPEX ratio of mining cost, we investigate alternative Proof of Work algorithms and complementary computing hardware paradigms that are difficult/expensive to produce but achieve high energy efficiency. One can observe that Artificial Intelligence (AI) hardware industry is converging to a similar goal as many companies try to commercialize exotic architectures for low-energy computing\footnote{Demand for AI compute is growing exponentially and cannot be supported by conventional hardware without massive energy consumption \cite{openai}.}. One of the promising approaches being commercialized for AI is optical computing, specifically photonic co-processors. Due to its commercialization feasibility and long term potential for ultra-low energy use, we concluded that optical computing is a promising platform for a low energy Proof of Work. 


\subsection{Optical Computing}

While in traditional digital hardware, we rely on electrical currents, optical computing uses light as the basis of its operations. The approach has been around for decades, however recent advances in the telecommunications industry and Artificial Intelligence (AI) have significantly contributed to optical computing development.
Indeed, researchers anticipate that integrating optical processing starting with on-chip signal routing and ending with 
optical accelerators for AI
can significantly boost processing speed, keeping energy consumption levels
as low as possible. Moreover, the semiconductor industry has nearly
reached its fundamental limits meaning that digital computers cannot continue improving on the same trajectory they have followed for fifty years \cite{Marr2013}.
This also means that continuous technological progress will require alternative computing methods and alternative hardware. The advantages of using light for information processing can be best illustrated by the adoption of optics in the telecommunications industry. Indeed, the replacement of copper cables with optical fibers transformed intercontinental communications, including the Internet, which has become exponentially faster, and more efficient.

Optical computing has a rich history dating back to Fourier processing in the 1940s (Duffieux 1946), the first optical neural networks in the 1980s (Psalti 1984), 1980s work on the optical transistor at Bell Labs, and modern work in optical neural networks, reservoir computing and optical quantum computing. Optical computing research at Bell Labs provided the inspiration for modern optical/photonic AI computing, whereas holographic computers and optical chaos communications influenced different flavors of optical reservoir computing.

\textit{Reservoir Computing} (RC) was initially conceived in the early 2000-s first as an attempt to simplify recurrent neural networks training \cite{Jaeger2001, Maass2002}.\footnote{A related method was proposed in cognitive neuroscience even earlier \cite{Dominey1995}.}
Trained in a supervised fashion by creating a simple linear readout from a high-dimensional state space, RC systems are capable of solving pattern recognition and time-series prediction challenges \cite{Ilies2007, Jalalvand2015} and quickly became popular for \textit{in-materio} computing with various physical substrates such as water (hence ``reservoir")\cite{Fernando2003}.  Researchers have shown that using optical systems for computation, photonic RC research can yield high-speed operation and low energy use. For comparison, an RC setup using off-the-shelf optical equipment performs speech recognition tasks about three times faster than a Google TPU \cite{Larger2017}. Another recent work implements photonic RC with a spatially-extended system \cite{Bueno2018}. This RC system, based on spatial light modulators and micromirror arrays, easily allows massively parallel information processing by hundreds of nonlinear nodes.

\textit{Deep Learning AI Computing} Although the original focus of brain-inspired computing was
to mimic the behavior of populations of biological neurons using integrated analog circuits, in recent years, photonic implementations of deep neural networks\footnote{DNN architectures do not mimic brain architecture directly, although they have kept the term ``neuron." Even though there exist neuromorphic photonic implementations that produce neuron-like integrate and fire processing\cite{neuromorphic}, they are not near commercialization.} have gained traction. As a result, several new optical hardware strategies have been developed to perform matrix-vector multiplication\cite{tait,shen}. Older approaches relied on slow spatial light modulators\cite{Cartwright1984}, whereas recent implementations,  which are now being commercialized, use integrated silicon photonic circuits based on microring resonators\cite{tait}, and Mach-Zehnder Interferometer arrays\cite{shen}. 

Despite successful demonstrations of optical computing in various forms over the years, the limiting factor for commercial adoption has always been the cost and difficulty of manufacturing the hardware on-scale. Competing digital computers have benefited from continuous improvements in fabrication technology and a fabless supply chain for new products\footnote{Companies bringing chips to market can work with foundries like TSMC and Global Foundries, rather than setting up their own manufacturing operations.}. The relatively recent advent of silicon photonics, which is compatible with standard chip manufacturing processes used for digital electronics, has opened the possibility of manufacturing scalable and reproducible optical co-processors. In the next two subsections, we provide a brief overview of silicon photonics and its application to deep learning AI processing. Additionally, in Section \ref{sec:optical-pow-prototype}, we briefly explain the working principles of matrix-vector multiplication in photonic circuits. 

\subsection{Silicon Photonics}  

\begin{figure}[h!]
    \centering
\includegraphics[width=0.46\textwidth]{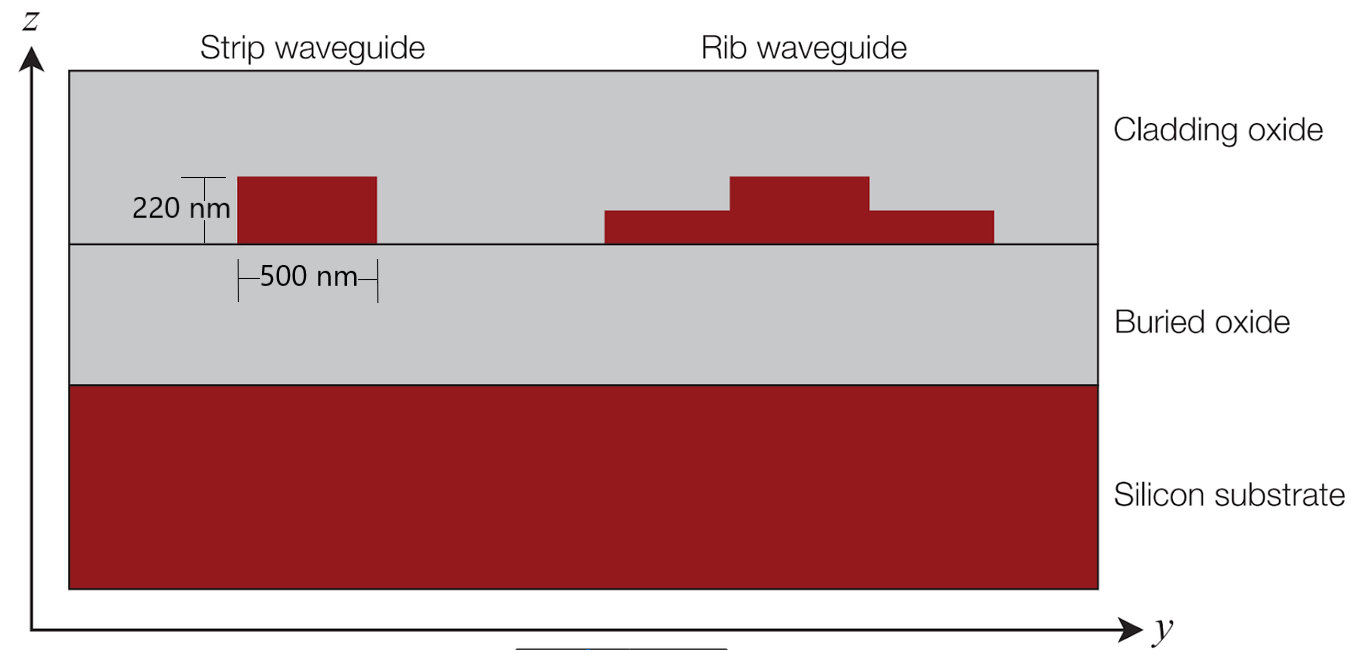}
  \caption{Cross Section of Silicon on Insulator Waveguides}
  \label{fig:SOI1}
\end{figure}

\begin{figure}[h!]
    \centering
\includegraphics[width=0.485\textwidth]{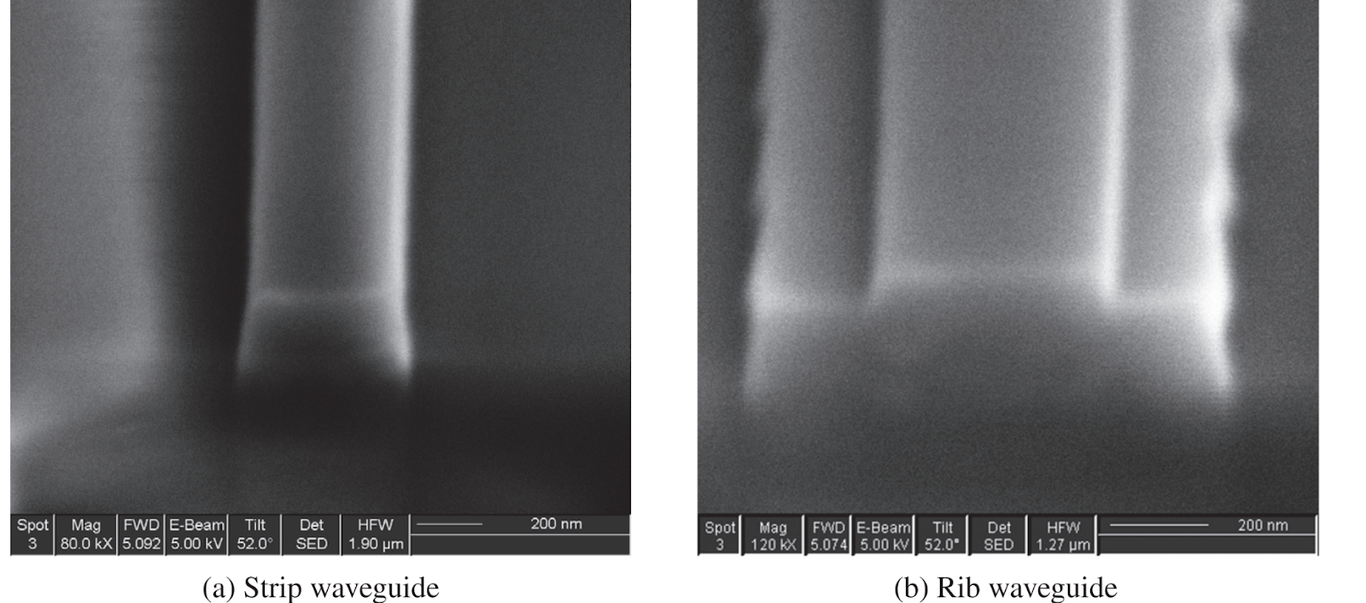}
  \caption{SEM Image of Silicon on Insulator Waveguides \cite{Silicon_Lukas}}
  \label{fig:SOI2}
\end{figure}

Traditionally, optical systems require precise alignment and expensive, carefully manufactured bulk components. Recent advances in Photonic Integrated Circuits (PICs) have addressed these problems successfully by porting bulk optical systems to chip-scale waveguide circuits. Photonic Integrated Circuits are produced by patterning thin dielectric or semiconductor wafers using micro/nano processing, leveraging the incredible and ever-increasing precision of lithography to ensure alignment and enable cheap mass production. Until the early 2000s, PICs were fabricated using expensive III-V materials. Despite silicon's inherently sub-optimal optical material properties (no commercial silicon lasers, no electro-optic effect), its ubiquitous use in electronics has created a huge fabrication ecosystem that made it advantageous to work around the problems and build on-chip silicon optical components. After breakthroughs in silicon (Si) photonic component design\footnote{See \url{https://youtu.be/csHshgggAdw} Michal Lipson's talk on the history of Si photonics} such as low loss optical fiber-to-chip couplers \cite{Fiber_Coupler}, fast electro-optic modulators\cite{Lipson_Modulator}, and Germanium on-chip photodetectors\cite{Lipson_Ge_Detectors}, it became possible to take advantage of the incredible developments in silicon CMOS\footnote{Complementary Metal–Oxide–Semiconductor (CMOS) is the nano-electronics fabrication process which has successfully decreased the cost and size of transistors by more than a factor of $10^6$ in last 60 years. The technology is the basis for modern digital circuits, including computer processors and memory.} technology over the last six decades to produce photonic circuits in re-purposed electronics foundry processes. One of the fundamental building blocks of Si photonic circuits is the nano-scale Si waveguide. Figures~\ref{fig:SOI1} and \ref{fig:SOI2} depict typical waveguides produced in a Silicon-on-Insulator wafer, which confine light within the photonic circuit via total internal reflection (the same effect used to guide light in fiber optics). 

Silicon photonic integrated circuits have had commercial success as transceivers for various datacom applications \cite{Semiconductor}. Today, millions of silicon photonic transceivers (manufactured by companies such as Luxtera, IBM, and Intel) shuttle information between server racks at data-centers. Additionally, there are now multiple companies commercializing silicon photonics for LIDAR and bio-sensing. Crucially, major microchip foundries (including Global Foundries and TSMC) are either offering Si photonics or are in the process of launching silicon photonic manufacturing lines. In recent years, this availability of commercial silicon photonic fabrication has spurred efforts to commercialize silicon photonic chips for massively parallel computation leading to the emergence of photonic co-processors for AI.

\subsubsection*{Integrated Photonic Co-Processors for AI}

Due to the recent success of deep learning AI algorithms, the demand for massive quantities of Multiply and Accumulate (MAC) processing has led to heavy investment in MAC processor research as well as many commercial efforts to produce specialized processors that perform these computations more efficiently from a cost and energy standpoint. In parallel to ongoing developments by GPU manufacturers like NVIDIA and Google (TPU), multiple companies such as Groq (digital), Graphcore (digital), as well as Mythic (analog) and Synthiant (analog) are pursuing innovative electronic architectures for MAC. 

AI architectures have been explored in free-space optical systems\cite{Saade2016, Bueno2018}, and several companies such as LightOn, Fathom, and Optalysis are working on implementing such systems commercially. However, more recently, due to the progress made by the Princeton Neuromorphic Photonics lab \cite{Prucnal} as well as research at MIT\cite{shen} and other academic institutions, several startups have emerged, including Lightelligence \cite{VentureBeat}, Lightmatter\cite{SPIE}, and Luminous\cite{Luminous}, that are applying silicon photonics designs for telecommunications and quantum information processing to build MAC processing photonic circuits. The promise of the technology, as detailed by Nahmias \textit{et al.}\cite{Mitchell}, is to offer 2-3 orders of magnitude better energy efficiency for MACs over electronic processors, and eventually even greater gains as optical computation has very high theoretical limits for energy efficiency\cite{Miller}. In a comparison of state-of-the-art GPU performance against a model of an electronic-photonic processor based on off-the-shelf foundry components, it was found a 2.8 to 14x speedup for the same power usage when performing CNN computations \cite{bangari2019digital}. An estimate for photonic co-Processors by Lima \textit{et al.} predicts 10fJ/MAC for a 128 channel chip vs. 1pJ/MAC for the Google TPU \cite{Thomas} and Nahmias \textit{et al.}\cite{Mitchell} predict that the performance can be pushed to 2.1fJ/MAC. These exciting developments in silicon photonic co-processors have created an opportunity for applying the underlying technology to low-energy applications outside AI processing. 

\subsection{Optical Proof of Work}

Inspired by the recent advances in silicon photonics for low-energy computation, we envision a practical PoW system built to complement optical computing. The main goal of such a PoW approach is to achieve drastic energy savings. Although, in the long run, it is conceivable that some miners will be built based on other analog architectures, we see photonic co-processors as holding the greatest potential for high energy-efficiency combined with near-term commercial availability. As a result, we propose Optical Proof of Work (oPoW), a PoW algorithm optimized for acceleration with integrated photonic co-Processors.


\section{Low Energy PoW}

Rather than attempting to compute an existing PoW algorithm using photonic hardware\footnote{There is enormous financial incentive to do this already—many optical computing experts have looked at the possibility of using photonics for bitcoin mining, however existing PoW algorithms are ill-suited to analog computing. Hashes like SHA256 are specifically designed to be efficiently implemented by digital processors.}, we chose to construct a modified PoW to favor existing photonic co-processor designs. Here we will briefly describe the co-design of a prototype photonic co-processor and PoW algorithm built to achieve our low-energy PoW goal.

It is worth noting that nearly all previous attempts to modify PoW algorithms to favor a specific hardware paradigm have focused on ASIC-resistance, meaning that rather than favoring specialized hardware, the aim is to exclude specialized hardware in favor of GPUs or CPUs\footnote{The end goal being democratization of hardware supply rather than energy efficiency or geographic decentralization.}. Examples include Scrypt, Cryptonight, Equihash, and, more recently, ProgPoW. Besides ProgPoW, which has not been implemented yet, these experiments have more or less failed due to the inherent advantages of specialized hardware. An excellent discussion of this topic can be found in \textit{The State of Cryptocurrency Mining}\cite{Vorick}, where the author concludes that: 

\begin{quote}
    ``For any algorithm, there will always be a path that custom hardware engineers can take to beat out general-purpose hardware. It’s a fundamental limitation of general-purpose hardware."
\end{quote} 

\noindent Optical PoW is fundamentally a simpler engineering problem than ASIC-resistant PoW. It is designed to be the most efficient on integrated photonics hardware, therefore giving one class of ASICs an advantage over another, rather than trying to limit the advantage of specialized hardware over general hardware.    


\subsection{HeavyHash}

Our goal in designing oPoW was to mimic the Bitcoin PoW construction (HashCash), maintaining the cryptographic security while ensuring that the PoW crypto puzzle is optimized for our “Target Paradigm” (photonic co-processors). As the major cost of PoW is evaluating the hash function of choice, the naive solution would be to find an optically computable hash.  However, a design choice was made early on to avoid all-optical hashes and Physical One Way Functions - due to issues of repeatability\cite{pappu2002physical} and their poorly understood security properties. Creating a new hash optimized for photonic processing was also not considered due to the complexity and risk of deploying an untested hash function, see IOTA fiasco\cite{heilman2017iota}. This leads to the selection of a hybrid design that composes digital hashing with low precision vector-matrix multiplication (intended for photonic acceleration) to produce \textit{HeavyHash}. HeavyHash is an iterated composition of an existing hash function, i.e.  SHA256, and a weighting function such that the cost of evaluation of HeavyHash is dominated by the computing of the weighting function. If the weighting function is dominated by the evaluation of a vector-matrix multiplication of sufficient size (and preferably unitary), it can be implemented with very high efficiency by photonic co-processors \cite{shen}. The ratio of the cost of computing the hashes versus that of the weighting function is tunable within a large range due to their different complexity orders of magnitude\footnote{An N increase in the output size of the hash, corresponding to an N increase in hash computation cost, leads to an N$^2$ increase in computation cost for the weighting function due to the properties of matrix multiplication.}).


\subsection{Optical PoW Prototype\label{sec:optical-pow-prototype}}
\begin{figure*}[ht]
    \centering
    \includegraphics[width=0.8\textwidth]{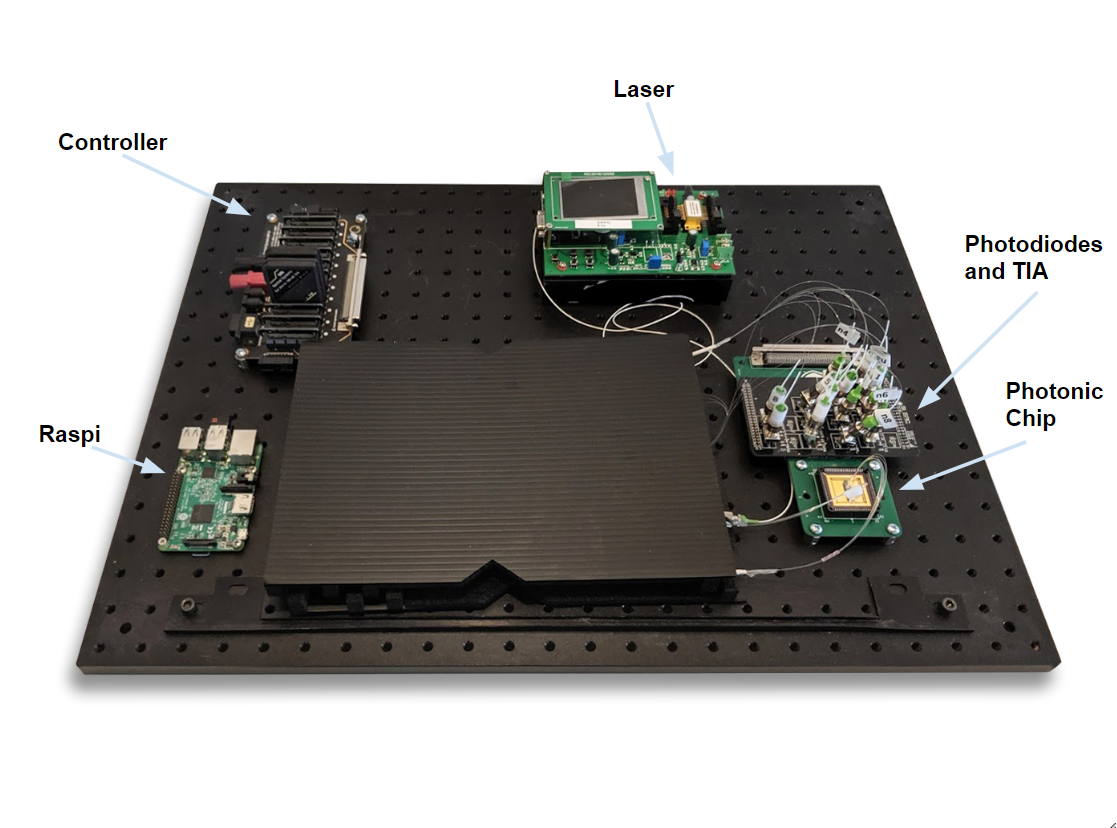}
    \caption{Prototype Miner Setup. Wiring excluded - the Raspberry Pi communicates with the Qontrol via USB, which in turn is connected to the PCB which carries the chip and TIA.}
    \label{fig:setup}
\end{figure*}

In its simplest form, oPoW is the Hashcash algorithm \cite{hashcash} with our custom hash function, HeavyHash, designed specifically to run efficiently on optical accelerators while preserving all PoW-necessary cryptographic security properties. A prototype hardware and software implementation of Optical PoW (an oPoW Bitcoin fork and a prototype oPoW silicon photonic miner) was developed with the goal of testing end-to-end functionality. Replacing the hash function in Bitcoin's PoW code is straightforward, so here we focus on the prototype hardware. Below is a short description of the system, the silicon photonic integrated circuit at its heart and the working principle of the analog computation.

\subsubsection*{System}

There are multiple known architectures for implementing an analog matrix-vector multiplier using standard silicon photonic components. The two main types of approaches are (1) the ring filter bank architecture developed at the Princeton neuromorphic photonics lab \cite{tait} and (2) various MZI interferometer meshes such as the triangular mesh used in the highly cited Shen \textit{et al.} \cite{shen} paper from MIT. 

Our analog photonic matrix-vector multiplier implementation is a rectangular directional coupler mesh\footnote{Mesh design was generated using an algorithm provided by Sunil Pai based on his work at Stanford\cite{pai2019matrix}.}. As seen in Figure~\ref{fig:setup} a RasPi board running our Bitcoin fork node software is paired with a driver board made by Qontrol, which communicates with a custom printed circuit board, TIA (to amplify the signal from the photodetectors) and interposer on which the silicon photonic chip is mounted\footnote{The PCB, TIA, and silicon photonic chip were fabricated in partnership with SiEPIC kits, an integrated photonics engineering firm affiliated with University of British Columbia.}. A close up of the packaged chip can be seen in Figure \ref{fig:chip_carrierw} and \ref{fig:chip} shows a top down view of the bare chip. The RasPi performs the digital portion of the HeavyHash and offloads the analog portion to the photonic chip via the Qontrol controller. 

\noindent
\subsubsection*{Silicon Photonic Chip}

Below in Figure~\ref{fig:chip}, is the layout of our prototype photonic chip. The chip consists of a single surface grating coupler input, split into 16 outputs. Each output is modulated (according to data provided by the RasPi) individually by a balanced thermal Mach-Zehnder Modulator\footnote{A brief explanation of the MZM is given in Section 2.2.3. For more details, an excellent tutorial on typical silicon photonics components can be found in \textit{Silicon Photonics Design: From Devices to Systems by Lukas Chrostowski and Michael Hochberg}\cite{Silicon_Lukas}}. The outputs of the modulators are then fed into the matrix multiplication directional coupler mesh\footnote{A brief explanation of directional couplers is given in 2.2.3}, and the outputs of the matrix multiplication network are collected into fibers via grating couplers and converted into electrical signals by the photodiodes and TIA. In a commercial setup, the photodetectors would be on-chip, making the entire package much simpler and eliminating the need for fiber optic connections (a laser input can be coupled via flip-chip bonding). 

\begin{figure}[h!]
    \centering
    \includegraphics[width=0.48\textwidth]{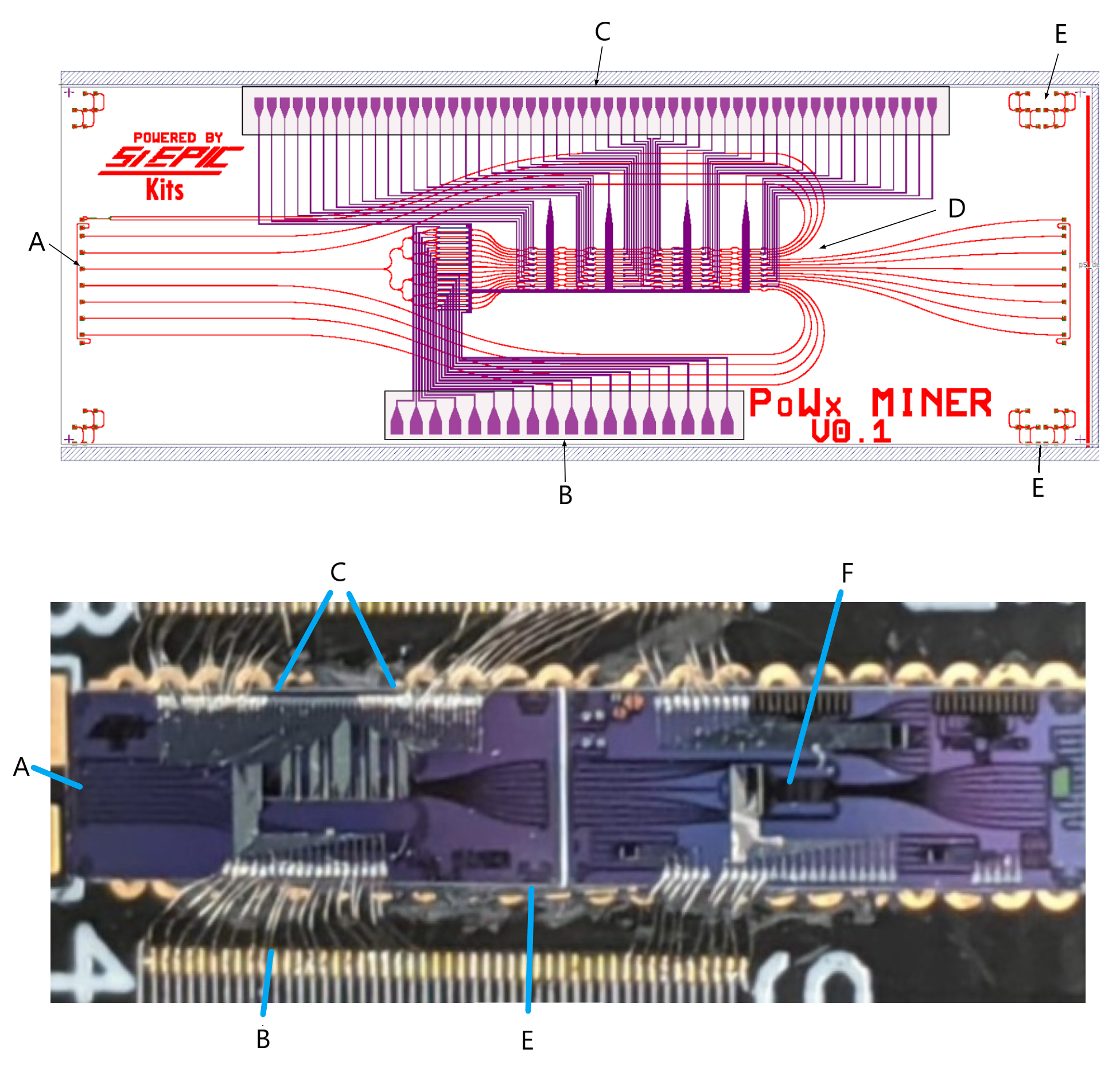}
    \caption{TOP: Photonic Circuit Diagram, A. Laser input (1550nm, common telecom wavelength) B. Metal pads for controlling modulators to transduce electrical data to optical C. Metal pads for tuning mesh of directional couplers D. Optical signal exits here containing the results of the computation and is output to fibers via a grating coupler the terminus of each waveguide. E. Alignment circuit for aligning fiber coupling stage.
    Bare oPoW miner prototype chip before wire and fiber bonding. On the left side of the die are test structures, oPoW miner is on the right. Bottom A-E: Same as TOP F. Test circuits}
    \label{fig:chip}
\end{figure}

\subsubsection*{Working Principle of Unitary Matrix Multiplication in a Mesh of Directional Couplers}

\begin{figure}[h!]
    \centering
    \includegraphics[width=0.48\textwidth]{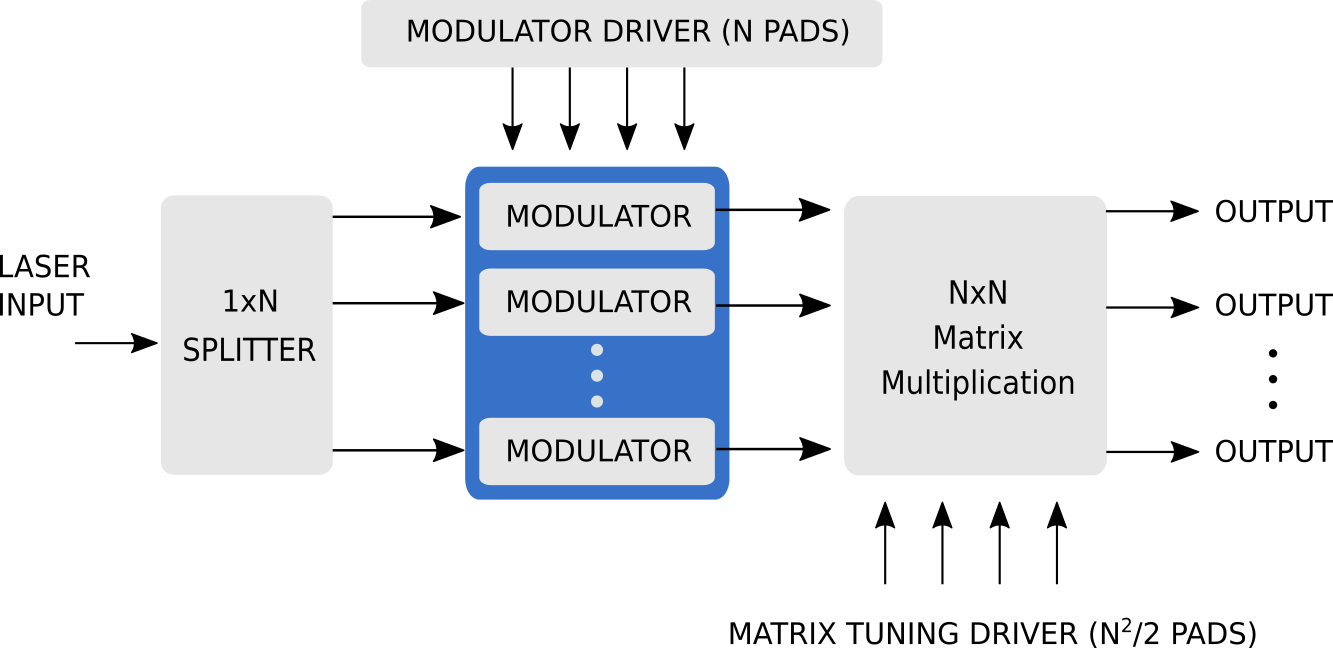}
  \caption{General Block diagram of the device's components (metal in blue). N = 16 in our design and the metal wirebonding pads provide electrical access to MZM modulators. A second set of pads provide access to tuning heaters in the directional coupler mesh.}
  \label{fig:Diagram}
\end{figure}

A generalized discussion of unitary matrix multiplication setups using photonics/interference can be found in Reck \textit{et al.} and Russell \textit{et al.} \cite{reck,russell}. In this section, we will provide a basic intuition of the working principle of the approach we used. 
\newline
\newline
As seen in Figure~\ref{fig:Diagram}, a single laser input is split evenly into multiple waveguides, each waveguide feeds into a modulator that can decrease the intensity of the light.

\begin{figure}[H]
    \centering
    \includegraphics[width=0.48\textwidth]{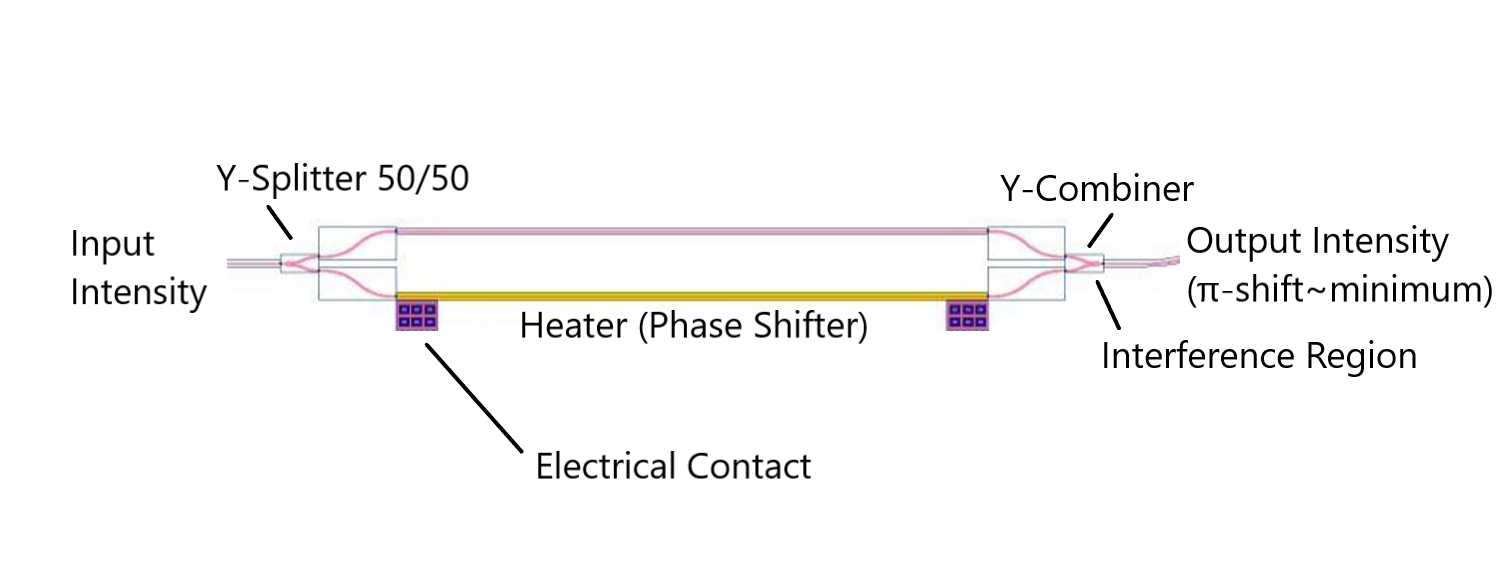}
  \caption{Mach-Zender Modulator}
  \label{fig:MZM}
\end{figure}

We chose a Mach-Zender Modulator, as seen in Figure~\ref{fig:MZM}, which splits the input light into two waveguides, and recombines them again with a phase shift. The phase shift is accomplished using a heater\footnote{There are PN junction-based phase shifters with much better speed and efficiency.} which changes the refractive index of one of the waveguides in the modulator.

In a perfect device, a full $\pi$ shift results in complete destructive interference, and smaller shifts can be used to get partial destructive interference. This is how a digital signal used to drive the modulator can be converted into an analog optical intensity. For example, in a 4-bit system, a $\pi$ shift would correspond to 0000, and zero phase shift would correspond to 1111, with partial interference providing the levels in between. The outputs of the modulators are then fed into a mesh of directional couplers (see Figure~\ref{fig:directional coupler} below), whose splitting ratios depend on the phases $\phi$ and $\phi'$ of the light entering at each input and the effective optical geometry of the coupling region\footnote{This depends on the physical geometry (length of coupler and the gap between the waveguides) as well as refractive index which can be tuned with a heating element).}.

\begin{figure}[H]
    \centering
    \includegraphics[width=0.48\textwidth]{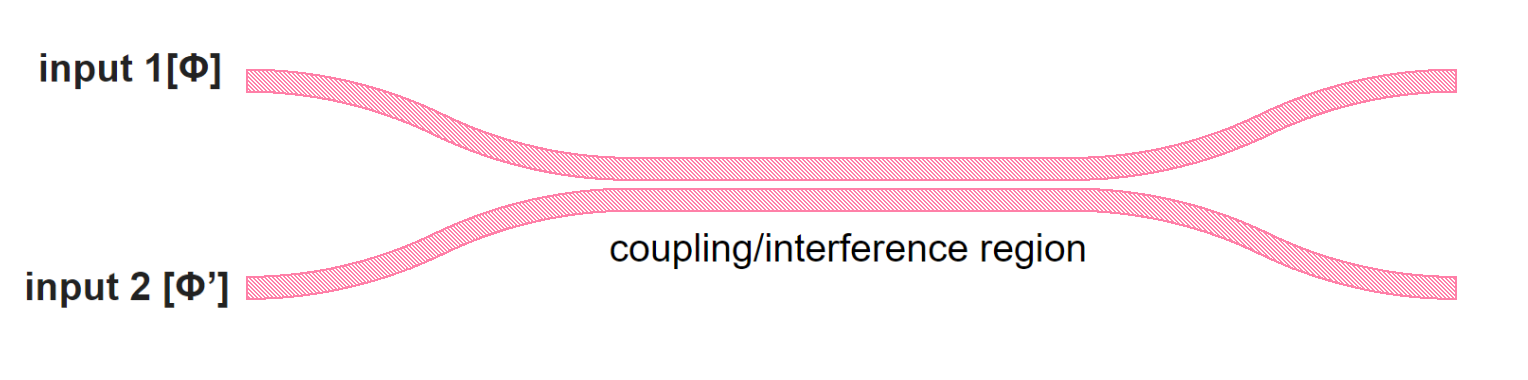}
  \caption{Basic directional coupler design}
  \label{fig:directional coupler}
\end{figure}


 By tuning the phase delays of each waveguide at each layer of the directional coupler mesh and the coupling region's effective optical length using heaters it should be possible to achieve an arbitrary unitary transfer matrix\footnote{Practically, our prototype can only achieve a subset of unitary matrices, due to limited number of electrical inputs but this is not a fundamental problem for commercial systems.}. If we call the vector of amplitudes and phases exiting each modulator the input vector [I]\footnote{By design the input phases are all the same.}, and the transfer matrix of the mesh [U], the output is mathematically equal to the matrix-vector operation [U]*[I]. However, the actual signal detected at the photodetectors corresponds to the intensity of light, not the phase. Therefore, the detected analog electrical signal
 actually returns the absolute value of the output vector [O]. Pai \textit{et al.} recently published a detailed discussion of this architecture and several other similar architectures with different trade-offs\cite{pai2019matrix}. 
 The advantage of using photonics to perform this operation, assuming a low precision AD-DA conversion is compatible with the use case, here is tremendous: all the light beams perform inherently parallel processing. 

We anticipate that adopting a PoW algorithm designed for photonic hardware will provide blockchain networks with various benefits, such as better security (higher 51\% attack resistance) and lower overall network energy consumption for equivalent networks. Although detailed analysis of the economics and security implications of a CAPEX-dominated PoW will be published separately, in the following sections we provide a brief summary.

\subsection{Energy Savings}

The total energy use of a PoW blockchain does not depend on the energy cost of a single hash/trial but on the total amount of energy used by the miners performing work in the system. For example, if SHA256 was replaced by a hash with lower computational difficulty (and therefore lower energy consumption per hash), but the same optimal hardware paradigm (ASICs), then miners would simply be forced to perform more hashes via difficulty adjustment. 

However, if the cost of PoW computation is biased towards a hardware paradigm which is more energy-efficient (but may be more expensive in terms of hardware cost), a blockchain built on such a PoW scheme will be more energy-efficient overall even if, somewhat paradoxically, an individual hash is more computationally expensive. 
The reason that this is possible is that the PoW difficulty adjusts so that the relative cost per block of the schemes is equivalent, even if the cost per single evaluation of the underlying hash function is significantly different. Because of the overall cost of the schemes depends on the value of the block reward, not the number of hashes required to get the reward, we can \emph{directly} compare the relative factors that make up the cost of each hashing scheme (energy cost, OPEX,  and hardware depreciation, CAPEX).\footnote{We are assuming here that the hashrate will be adjusted by market forces so that the relative cost of mining is identical in either framework.}

So long as the more energy-efficient hardware paradigm provides some marginal cost advantage over others, we assume that rational agents will adopt this paradigm to maximize their utility in a given mining ecosystem.
In summary, a low energy PoW can be achieved by tailoring a PoW algorithm to a hardware paradigm with a CAPEX dominated cost per hash/trial.  

\subsection{Security Budget Implications}\label{sec:51}

The Bitcoin network pays miners over \$5B yearly to secure its ledger. In the end, this cost is borne by the holders of Bitcoin via inflation. When analyzing any proposed security/consensus algorithm for decentralized cryptocurrencies, the key question is: \textit{How much real security does the security budget buy?} 
\newline

\noindent To examine the economic security of implementing oPoW in a blockchain protocol, we consider the classic $51\%$ attack\cite{Bitcoin}, as well as hashrate behavior over time.

\subsubsection{Fifty-One Percent Attack Security in a Low-OPEX PoW}

 In the 51\% attack, an adversary is interested in acquiring more than half of the \textit{hash-power} of the system in order to break consensus, build their own parallel chain and do things like double-spend payments or censor specific payments by rewriting the blockchain history within their secret chain. The cost of such an attack would be to match (and surpass) the total CAPEX (hardware controlled by honest nodes) in the system and pay for the OPEX (energy) cost for the duration of the attack. Any PoW blockchain system's security is predicated on a high cost for such an attack. 

Assuming a single system using a particular implementation of the oPoW algorithm\footnote{The analysis gets complicated if multiple networks are using the same PoW algorithm, because miners from one can attack the other.}, an attacker willing to acquire $51\%$ of the hashrate for an attack likely can't rent the hardware necessary for this attack. Miners in the system are unlikely to simultaneously rent such a large portion of their hashrate\footnote{Although not impossible, a coordinated effort to rent on the order of half of the hardware on a network would be difficult to hide and owners of hardware have an incentive not to rent to attackers, as the hardware is likely to lose value if the network is attacked.}, and since no other system is using the hardware there is no secondary source (a system with generic hardware, such as GPUs, doesn't have this advantage). Thus the attacker must purchase close to the total CAPEX of the system to gain 51\% of the computing power. Note that by attacking the system, the attacker potentially makes the resale value of the hardware negligible.

This analysis also holds true for Bitcoin. Although Bitcoin has high OPEX, the cost of a short attack (on the order of days/weeks) is dominated by the cost of acquiring the necessary hardware. Overall, oPoW actually provides greater 51\% security than an OPEX-heavy PoW in the long run, as it leads to faster hashrate growth and greater hashrate resilience to decreases in the value of the block rewards.\footnote{Note that in some cases energy cost could be partially CAPEX rather than fully OPEX if the miner is actually investing in infrastructure. Certainly this is better in terms of hashrate stability as long as the infrastructure cannot be re-purposed, however energy infrastructure is bulky and immobile compared to computing hardware and requires more upkeep.} 

\subsubsection{Hashrate Growth and Resilience in a Low-OPEX PoW}

Shifting mining cost from OPEX to CAPEX, increases the total effective investment made by the network (via block rewards and transaction fees) into long term security. Any OPEX costs the miners incur do not contribute to the hashrate growth and therefore, the long term security. OPEX is a necessary evil. As more funds flow to CAPEX, the network builds up a larger and larger cache of specialized security hardware, making the barrier for attack higher. In a related positive effect of CAPEX dominance, miners running low-OPEX hardware have no reason to turn it off when the coin price (and therefore mining reward value) or the price of electricity fluctuates. Bitcoin's hashrate growth is not nearly as impressive as it looks when hardware performance improvements are accounted for. Analyzing a more nuanced metric, perhaps\textit{Specific Hashrate} (we can loosely define Specific Hashrate as hashrate divided by the dollar cost of performing a single hash), shows that Bitcoin's security is very sensitive to price. In Q4 2018, Bitcoin prices were volatile, and the coin temporarily lost around 45\% of its market value. As a result of miners shutting off their machines to avoid paying for electricity, the hashrate dropped from 60 EH/s to 35 EH/s\cite{Bitinfo} (despite Bitmain releasing a new high performance 7nm miner\cite{huillet_2019} and other hardware manufacturers joining the fray). oPoW's economics can create a faster-growing, more stable, and more committed community of miners.  

\section{Considerations for a Practical Implementation of oPoW}

\noindent Below, we discuss some of the basic factors required for successful real-world oPoW implementation. We do not intend this list to be exhaustive, but instead, highlight some of the key considerations.

\subsection{HeavyHash Security Properties}

HeavyHash draws its cryptographic security from the hash (such as SHA256), which is composed with the weighting function to create the HeavyHash construction. As long as the weighting function preserves the entropy of the initial hash output, the random oracle security properties of the hash are inherited. In order to give photonic co-processors an advantage in evaluating HeavyHash, the weighting function must not only be optimized for photonic co-processors, but it must be digitally hard across most instances such that there is no better algorithm for evaluating the entire HeavyHash in digital ASICs than straightforwardly computing the entire HeavyHash for each new trial/nonce. We call this concept \textit{Minimal Effective Hardness}. Beyond these intuitions, the specifics of the algorithm and a detailed proof of its security will be published in a separate manuscript.~\cite{oPoWpaper}

\subsection{Energy Savings}

For an oPoW implementation to deliver the drastic energy savings relative to equivalent conventional Proof of Work blockchains, two key requirements will need to be satisfied:

\begin{enumerate}
    \item Miners using photonic co-processors must have a lower total cost per hash (amortized CAPEX + OPEX) than competing hardware (i.e. ASICs, GPUs).
    \item The ratio of CAPEX to OPEX in the cost per hash for photonic oPoW miners must be an order of magnitude higher than it is for ASICs and GPUs currently running on networks like Bitcoin and Ethereum. 
\end{enumerate}

\noindent Based on internal engineering and extensive discussions with researchers and hardware companies working in photonic computing, we believe that these are achievable goals given the state of the art today, however, a live implementation of oPoW will provide an empirical test. 

\subsection{Decentralization}

Supposing drastic energy savings are achieved, it can be argued that mining decentralization would necessarily be a direct result. Below we briefly discuss two aspects of the issue.  

\subsubsection{Geographic Decentralization}

Although there will always be a small energy cost, and therefore some kind of savings associated with operating in cheap-energy regions, energy will no longer be the deciding factor in profitability. There is a pent-up demand for mining participation in big cities and other areas with expensive energy but crypto-friendly laws (i.e., Malta). Currently, potential miners in these areas have no way to get a return on capital because their operating costs would be higher than the value of the cryptocurrency rewards they would be able to capture via mining. oPoW will democratize mining and provide miners an opportunity to operate in more crypto-friendly jurisdictions with lower risks, the rule of law, and lower costs of capital.

We expect to see big miners in low-energy regions continue mining conventional PoW coins, where they will have less competition, as new players emerge in other regions to mine on oPoW networks. 

\subsubsection{Hardware Manufacturing Decentralization}

In addition to energy efficiency, silicon photonics as a platform has the advantage of lower NREs (non-recurring engineering expenses) as silicon photonic circuits are fabricated using older process nodes (i.e., 200nm SOI\cite{AIM}, 90nm SOI\cite{Global_Foundries} vs. 7nm for Bitcoin ASICs\cite{Bitcoin_News}). Low NREs will work to decrease barriers to entry and ensure a healthy, competitive supplier market for oPoW miners in the long run. Additionally because oPoW is based on a photonic co-processor architecture that is being applied more generally to AI processing, we expect there to be robust supplier competition in oPoW mining hardware. Not only are there multiple companies commercializing AI photonic co-processors as discussed in the introduction, but there are also other approaches to analog matrix-vector multiplication being investigated, such as crossbar memristor arrays and other electronic brain-inspired architectures which could eventually deliver competing miners to the market. A broader intuition worth mentioning: it is much easier for a single manufacturer to dominate the market for a hash like SHA256 that has no high-performance computing use-case outside cryptocurrency than it is for a single manufacturer to dominate the market for computing a more general operation that is used beyond a particular coin's PoW. 

\section{Long Term Outlook}
\smallskip

\noindent \textit{Cryptocurrency} Cryptocurrencies have progressed in the past five years from the concept stage to early commercialization. It is hard to estimate the true long term potential of the technology, however, it is clear that there is an opportunity to increase the efficiency and fairness of the global financial system. Access to cryptocurrency markets can act as a safety valve in crisis situations, and we have already seen this happen as fiat currency crashes in countries like Zimbabwe and Venezuela have led to local spikes in Bitcoin demand. As cryptocurrencies become more stable, functional and user-friendly they will be able to compete with traditional financial services more broadly.


\smallskip

\noindent \textit{Silicon Photonic Co-Processors} There is a lot of hope in the photonics industry that the success of silicon photonics in data communications will translate to computational use cases. It is clear that computing with photons instead of electrons offers attractive fundamental advantages, however, numerous practical engineering challenges must be met to apply optical computing broadly. Taking advantage of the standard semiconductor fabrication supply chain by using silicon photonics addresses many of the major problems, however there are some still remaining, such as the absence of a silicon laser source\footnote{There has been a lot of success with hybrid silicon III-V lasers \cite{Ge_Laser,bowers} as well as more exotic attempts at silicon lasers \cite{Si_Laser,Raman_laser}.}. In the case of AI processing, suitable optical nonlinear components that are compatible with semiconductor foundries are an active area of research, however not yet commercially available. Additionally, components in silicon photonic circuits are typically tuned to adjust for manufacturing variance using microheaters which, leads to an increase in the overall energy consumption of any photonic circuit\footnote{There is a lot of interesting work being done to replace microheaters, such as non-volatile phase change materials tuning \cite{JJ}.}. We anticipate that the simplicity of the Optical Proof of Work use case (brute force computation designed for photonics, with nearly no memory requirements or variability) will prove to be an excellent stepping stone for photonic co-processor technology on it's way to mainstream commercialization.

\smallskip
\
\noindent \textit{Optical PoW} Scaling store-of-value cryptocurrencies, Bitcoin and others, to meet global demand will require both technical and social innovations. Numerous researchers and developers are working to make improvements via off-chain developments such as the Lightning Network\footnote{Lightning is a framework for increasing transactions processed per second for the Bitcoin network via off-chain payment channels that eventually settle their final totals to the Bitcoin blockchain} and fundamental blockchain innovations such as MimbleWimble / Zcash / Monero (privacy), and DAGs (scalability). Entrepreneurs are improving new-user onboarding and generally smoothing out the experience for non-technical users. However, besides interesting efforts (albeit very centralized) to use renewable energy for Bitcoin mining, Proof of Work has not seen much innovation since the advent of Bitcoin mining ASICs in 2012.


Our goal at PoWx is to change that by taking advantage of next-generation computing. A fundamental shift in the PoW ecosystem is needed to support another order-of-magnitude increase in decentralized store-of-value. While requiring minimal modifications to existing Proof of Work schemes and thus inheriting desirable security properties, Optical Proof of Work has the potential to solve some of the deepest issues faced by Bitcoin and other cryptocurrencies today. oPoW has the promise to untether cryptocurrencies from power plants, enabling geographically decentralized mining and therefore improved security with additional benefits of eliminating the sensitivity of hashrate to coin price, and democratizing issuance. The implementation of oPoW will help accelerate the development of energy-efficient photonic co-processors, acting as a stepping stone to other applications.

\section*{Acknowledgements}

The authors acknowledge partial financial support from PoWx. 
We thank Sunil Pai (Stanford) for assistance in prototype chip design and Mustafa Hamood, Stephen Lin and Jaspreet Joha (SiEPIC/University of British Columbia) for assistance in prototype fabrication and testing. Additionally we would like to thank Guy Corem (Beam), Bram Cohen (Chia), Tom Brand (Starkware), Yichen Shen (Lightelligence), Mitchell Nahmias (Luminous Computing), Yonatan Sampolinsky (DAGlabs), and John Tromp (Grin) for helpful discussions and feedback.

\onecolumn{

\begin{flushleft}

\bibliographystyle{unsrt}

\end{flushleft}
}

\end{document}